\documentclass[twoside,a4paper]{article}
\usepackage[T2A]{fontenc}
\usepackage[cp1251]{inputenc}
\usepackage[english]{babel}
\usepackage{graphicx}
\makeatletter
\renewcommand{\@cite}[2]{{#1 \if@tempswa   #2\fi}}
\renewcommand{\@biblabel}[1]{\hfill}
\makeatother

\newcommand{\beq}{\begin{equation}}
\newcommand{\eeq}{\end{equation}}
\newcommand{\bea}{\begin{eqnarray}}
\newcommand{\eea}{\end{eqnarray}}

\def\sun{\hbox{$_\odot$}}
\def\arcsec{\hbox{$.\!\!^{\prime\prime}$}}
\def\degree{\hbox{$.\!\!^{\circ}$}}

\begin{document}

\vbox{}

\begin{center}
{\LARGE  \bf Deceleration of SS\,433 radio jets}
\vspace{0.7cm}

{\large  A.~A. Panferov}

IMFIT, Togliatti State University, Russia

E-mail: panfS@yandex.ru
\end{center}
\vspace{0.7cm}

\begin{abstract}
The mildly relativistic jets of SS\,433 are believed to inflate the surrounding supernova 
remnant W\,50 depositing in its expansion possibly more than 99\% of their kinetic energy
(\cite[1998]{Dub98}). 
Where and how this transformation of  energy is curried out, it is not yet known. 
What can we learn from it that the jets decelerate and the deceleration is non-dissipative, 
i.e. radiatively dark. 
In this paper we unclose the observed deviations of the precessing radio jets of SS\,433,
within a few arcseconds from a jets source, from the ballistic track, described by the kinematic 
model, as a signature of the deceleration which, on other hand, issues from the jets
colliding with ambient medium. 
For that we model kinematics of these colliding jets. The ram pressure on the jets is estimated from
the observed profile of brightness of synchrotron radiation along the radio jets. We have found
that to fit observed locus the radio jets should be decelerated and twisted, additionally to 
the precession twist, mostly within the first one-fifth of precession period, and further they 
extend imitating ballistic jets. 
The fitted physical parameters of the jet model turned out to be physically reliable and characteristic
for SS\,433 jets that unlikely to be occasional. This model explains naturally, and meets approval by
a) the observed shock-pressed morphology of the radio jets and their brightness,
b) the observed $\sim 10\%$ deflections from the standard kinematic model --- 
just a magnitude of the jet speed decrement in the model, 
c) regularly observed for the radio jets the precession phase deviation from the standard
kinematic model prediction, 
d) dichotomy of distance to the object, 4.8\,kpc vs. 5.5\,kpc, 
determined on the basis of the radio jets kinematics on scales of a sub-arcsecond and several 
arcseconds. The latter fact once again convinces that the kinematic distance to SS\,433
could not be correct without accounting for the jets deceleration. 

The model proposed here several reveals evolution of SS\,433 radio jets with distance.

\end{abstract}

\section{\large Introduction}
Jets of SS\,433 are the brilliant representative of star relativistic jets (see review of
\cite[2004]{Fab04}). In radio wavelengths the jets are viewed from some milliarcseconds to some 
arcseconds as a corkscrew --- a signature of jets precession. 
Kinematics of the radio jets as well as the shifts of spectral lines of the X-ray (at the distances 
from a jets source $z\sim 10^{11}$\,cm) and optical (at the distances $z\sim 10^{15}$\,cm) 
jets fulfill
the same (standard) kinematic model, according to which each element of the jets moves freely 
(so-called ballistic movement) along the jets axis, which precesses and nutates
(\cite[1979]{AM79}; \cite[1981]{HJ81}; \cite[1981]{NC81}). Though the radio jets in the
beginning are intimately connected with the optical jets, which are believed to be ballistic 
(\cite[1987]{Kop87}), its behaviour some differs: the radio jets departure from kinematics of the
optical jets by as much as 10\% (e.g. \cite[2004]{BB04}; \cite[2004]{Sch04}; \cite[2008]{Rob08}); 
the radio clouds could be found far out of jet axis (e.g. \cite[1984]{Sp84}; \cite[1987]{Rom87}); 
the radio jets appear more continuous vs. the bullet-like optical jets (\cite[1987]{Bor87});
the radio jets are outstanding for the zones of a qualitative change in jet flow at distances of
$0\arcsec050 - 0\arcsec100$ (the radio brightening zone, \cite[1987]{Ver87}) and 
$\sim 1\arcsec5$ (the reheating zone, observed in X-ray, \cite[2002]{Mig02}). 
These peculiarities could be due to the dynamics of the jets.
In the environments of SS\,433, with the powerful wind of a mass flow rate $\sim 10^{-4}$~M\sun/yr 
and a velocity $v_{\rm w} \sim 1500$~km/s from supercritical accretion disk, the wind-jet
ram pressure might radically influence hehaviour of the precessing jets if there was not anisotropy
of the wind nor evacuation of gas from the jets channel by previous precession runs of the jets.

Nevertheless the jets should decelerate somewhere on a way to the shell of supernova remnant
W\,50 encircling SS\,433: the flight time of the jets with the velocity $v_{\rm j}\approx 0.26\,c$ of 
the optical jets, where $c$ is the speed of light,
over W\,50 radius $R_{\rm W50}/v_{\rm j} \approx 80\,{\rm pc}/0.26\,c \sim 1000$~years
is much smaller than the age $\sim 10^4$~years of W\,50. Moreover, on the basis of the X-ray data
the jets lose their helical appearance and are decelerated already in interiority of W\,50
(\cite[2007]{Bri07}).  And at the same time the jets are dark: only less than 1\% of the huge
kinetic luminosity $L_{\rm k} \sim 10^{39}$~erg/s of the jets goes into jet radiation and
heating of W\,50 and the rest likely goes into mechanical energy of W\,50, with an unbelievable
effectiveness of $\sim 99$\% (\cite[1998]{Dub98}; \cite[2007]{Bri07}). 

Other indications of the deceleration may be the followings: 

1. The difference of jet speed in X-ray, at distance $z\sim 10^{11}$~cm, and optical,
at $z\sim 10^{15}$~cm, spectral domains --- $0.2699\pm0.0007\,c$ (\cite[2002]{Mar02}) vs.
$0.2581\pm0.0008\,c$ (\cite[2008]{Dav08}). This corresponds to the rate of loss of
jet kinetic energy $\dot{W}_{\rm k}/{L}_{\rm k} \approx 2 \delta v_{\rm j} = 9\%$,
if the jet mass flow rate $\dot{M}_{\rm j}$ is constant.

2. As \cite[(1987)]{Kundt} has already remarked the kinematic distance to the object determined 
by kinematics of the extended jets (on a scale of several arcseconds), 
is more than the distance determined by kinematics of the inner jets (on a scale of sub-arcsecond):
$5.5\pm0.2$~kpc (\cite[1981]{HJ81}; \cite[2004]{BB04}) from
unresolved in time images of the extended jets vs. the estimations from inner jets 
$4.9\pm0.2$~kpc (\cite[1984]{Sp84}), $4.85\pm0.2$~kpc (\cite[1993]{Ver93}) and
$4.61\pm0.35$~kpc (\cite[2002]{St02}) from proper motion of radio knots, 
and $5.0\pm0.3$~kpc (\cite[1986]{Fejes86}) and $5.0\pm0.5$~kpc (\cite[1987]{Rom87}) 
from unresolved in time images. This dichotomy could be explained by the fact that 
a priori assumption of the constancy of the speed $v_{\rm j}$ in kinematic simulation 
of really decelerated extended jets leads to an overestimation of the distance $D$ to SS\,433 to meet 
the observed angular size, which in the first approximation is proportional to the ratio 
$v_{\rm j} /D$. So \cite[(2008)]{Rob08} note that if to accept for the extended jets 
the kinematics with the set $v_{\rm j} /D = 0.2647\,c /5.5$\,kpc, the kinematics of the 
inner jets agrees better with the set $0.2647\,c /5.0$\,kpc or with the velocity greater 
than that of the optical jets while holding the distance, $0.29\,c /5.5$\,кпк. In this regard, 
they suggest the variations of the radio jet velocity $\pm10\%$.
 
3. The velocity profile along the radio jets received by \cite[(2004, Рис.4)]{BB04}, 
shows variations of $v_{\rm j}$ with distance along the jets. Scope of these variations corresponds 
to the rate of jet speed change $\le 0.018\,c/P_0$, where $P_0$ is the precession period.

\cite[(2004)]{St04} found that jet model fit to the inner and extended radio jets on the same image 
is much better when the jet kinematic model with the distance $D = 4.8$\,kpc enables the
deceleration $0.02\,c/P_0$. What \cite[(2011)]{Bell11} noticed that a satisfactory fit can be obtained 
with a constant speed, if to use a larger distance, $D = 5.5$\,кпк, that  
supports estimations of the same distance in (\cite[2004]{BB04}; \cite[2008]{Lock08}).
However, the choice of constant $v_{\rm j}$ and $D = 5.5$\,kpc does not eliminate the 
above distance dichotomy, resulting from many observations.

Recently was received the brightness profile of synchrotron radio emission along the jets of SS\,433 
up to a distance of 800 days of flight (\cite[2010]{Rob10}; \cite[2011]{Bell11}).
Emitting relativistic electrons there are probably accelerated in shock waves resulted 
from the colliding of the jets with environments (\cite[1990]{Hea90}; \cite[2002]{Par02}). 
This profile allows to evaluate the dynamic pressure of the medium on the jets and, 
as a result, their deceleration. In our work we use this profile for research of the model 
of the radio jets that satisfies the observed kinematics both internal and external jets 
at the same distance to an observer.  

\section{\large Dynamics of SS\,433 jets}
The morphology of the radio jets on the maps of \cite[(2008)]{Rob08} gives a clear indication 
of a significant role in its formation of dynamic pressure of the environments.
We propose a model in which the jet is quasi continuous and clumped, 
and the gas from environment  swept by jet provides dynamic pressure 
on the jet surface, resulting in deviation of the jet from ballistic kinematics.
The shock waves arising on the surface and spreading inside the jet can cause strong turbulence
at density inhomogeneities in the jet and, as a result, strengthen the magnetic field, and 
very effectively accelerate the relativistic particles, as it is in supernova remnants 
(\cite[2012]{Ino12}). In contrast to the model of \cite[(1988)]{HJ88} in our model 
the synchrotron radiation of the radio jets occurs not on the surface of homogeneous 
jets, but at the vicinity of the dense clouds, and adiabatic jets are not expected:
the radio jets heat up to temperature $T > 10^7$~K (\cite[2002]{Mig02}).

According to this model, the average internal jet pressure $p_{\rm in}$, equal to the sum of the
pressures of magnetic field $p_{\rm m}$, relativistic particles $p_{\rm r}$ and gas $p_{\rm g}$, 
should be approximately equal to the dynamic pressure $p_{\rm dyn}$ on the surface of the jet,
neglecting the magnetic field and gas pressure of the environment --- we believe their equality: 
\beq
p_{\rm dyn} = p_{\rm in} \equiv p_{\rm m}(1+\frac{1}{3 \beta_{\rm H}} + \beta_{\rm g}),
\label{press_d}
\eeq
where $\beta_{\rm H} = \epsilon_{\rm H}/\epsilon_{\rm r}$ is the ratio of energies of the 
magnetic field $\epsilon_{\rm H}$ and of the relativistic particles $\epsilon_{\rm r}$, 
$\beta_{\rm g} = p_{\rm g}/p_{\rm m}$.
Profile of the magnetic pressure of $p_{\rm m}$ along the jet is evaluated using the spectral 
density of observed synchrotron radiation flux $S_\nu$ (hereinafter brightness) in jet comoving
frame of reference, in the assumption of power-law energy spectrum of electrons 
(e.g \cite[1979]{Gin79}): 
\beq
p_{\rm m} \equiv H^2/8\pi=(k_\nu \beta_{\rm H}\beta_{\rm e} D^2 S_\nu/V)^{4/7},
\label{press_m}
\eeq
where $\beta_{\rm e} = \epsilon_{\rm r}/\epsilon_{\rm e}$ is the ratio of the energy of 
relativistic particles to the energy of relativistic electrons $\epsilon_{\rm e}$, $H$ the magnetic 
field strength, $V$ the volume of the radiating gas within a beam of telescope.  
In the case of SS\,433 coefficient $k_\nu$ equals to $5.47\cdot 10^{17}$ 
for a synchrotron spectral index 
$\alpha=0.74$ ($S_\nu \propto \nu^{-\alpha}$), the radiation frequency $\nu = 4.86$~GHz,
at which \cite[(2011)]{Bell11} defined the brightness $S_\nu(t)$ as a function of flight 
time $t$, and for the frequency range $4.86 \div \infty$~GHz accepted as 
a whole power-law radiation spectrum of the relativistic electrons. The pressure $p_{\rm m}$ may 
be underestimated maximum $2\div3$ times due to uncertainties in the lower border of the 
synchrotron spectrum.

In our model, the magnetic field inside the jets and, therefore, the region of synchrotron 
radiation is localized at the vicinity of dense clouds, in the shell of the thickness $l_{\rm sh}$ 
approximately equal to the size of clouds $l_{\rm cl}$, as \cite[(2012)]{Ino12} demonstrated 
in the case of clouds in supernova remnants. For a jet segment of unit length
equation (\ref{press_m}) is transformed to
\beq
p_{\rm m}=(k_\nu (1+\frac{1}{3 \beta_{\rm H}} + \beta_{\rm g}) 
\beta_{\rm H} \beta_{\rm e} \mu m_{\rm p} 
\lambda_{\rm j} D^2 s_\nu/ k_{\rm V} k_{\rm B} \dot{M}_{\rm j} T_{\rm cl})^{4/3}
\label{press_r}
\eeq
аfter substituting in equation (\ref{press_m}) the volume of the clouds shell
\beq
V= \frac{k_{\rm V}}{\rho_{\rm cl}} \frac{\dot{M}_{\rm j}}{\lambda_{\rm j}} =
\frac{k_{\rm V} k_{\rm B} T_{\rm cl}}{\mu m_{\rm p} (1+\frac{1}{3 \beta_{\rm H}} + \beta_{\rm g}) 
p_{\rm m}} \frac{\dot{M}_{\rm j}}{\lambda_{\rm j}},
\label{vol}
\eeq
and resolving obtained expression with respect to $p_{\rm m}$.
Here $s_\nu = {\rm d}S_\nu/{\rm d}l$ is the differential spectral density of the 
synchrotron radiation flux, or brightness per a jet unit length, $l$ the jet length, 
$k_{\rm V}$ the ratio of the volumes of cloud shell and cloud itself, 
$k_{\rm V} = (2l_{\rm sh}/l_{\rm cl}+1)^3-1$ in the case of a spherical cloud,
 $\rho_{\rm cl}$ and $T_{\rm cl}$ the mass density and temperature of the cloud,
which gas pressure $p_{\rm cl}=\rho_{\rm cl} k_{\rm B} T_{\rm cl}/\mu m_{\rm p}$
is assumed dominant and equal to the pressure $p_{\rm in}$ of intercloud medium,
$\dot{M}_{\rm j}$ the rate of jet mass flux contained in the clouds,  
$\lambda_{\rm j}$ the jet length per unit of the flight time, otherwise 
the jet speed in the frame of reference corotating with jet, 
$ \mu$ the average relative molar mass of clouds ($\approx 0.6$ for solar abundances), 
$m_{\rm p}$ the proton mass, $k_{\rm B}$ the Boltzmann constant.

Temperature of the clouds presumably increases with the distance $z$ from a jets source, 
in line with the observed X-ray brightening of the jets (\cite[2002]{Mig02}). In our model 
the power-law profile $T_{\rm cl}(z) = T_0 z^{\rm n}$ was adopted. 
The following hence model of clouds evolution, given the profiles of pressure 
and temperature, is limited by a condition on volume filling of the jets by the clouds:
\beq
f(z)=\frac{\dot{M}_{\rm j}}{\rho_{\rm cl}} \frac{1}{\pi (z\theta_{\rm jr}/2)^2 v_{\rm j}} 
= \frac{k_{\rm B} T_{\rm cl}\dot{M}_{\rm j}}{\mu m_{\rm p} 
(1+\frac{1}{3 \beta_{\rm H}} + \beta_{\rm g}) p_{\rm m}}
\frac{1}{\pi (z\theta_{\rm jr}/2)^2 v_{\rm j}} < 1.
\label{fil}
\eeq
Here it is assumed a conical geometry of the radio jets, with the openning $\theta_{\rm jr}$, 
and the volume of a jet segment is approximated by the cylinder volume
$\pi (z\theta_{\rm jr}/2)^2 v_{\rm j}$ since the expansion of the jets is negligible: $v_{\rm j}/z\ll1$.

The extended radio jets are not showing the nutation pattern, although there are the signs of  nutation 
of the inner radio jets within the zone of brightening (\cite[1993]{Ver93}, \cite[2007]{Miod07}). 
This allows to suggest that the nutation structure is blurring in the radio jets, and they
acquire an opening $\theta_{\rm jr} = \theta_{\rm j} + 2\theta_{\rm n} = 6\degree8$, where
$\theta_{\rm j} = 1\degree2$ is the opening and $\theta_{\rm n} = 2\degree8$ 
the nutation cone semi-opening of the optical jets (\cite[1987]{Bor87}).
In this case the angular velocity of the radio jets is the precession one, and the length
$\lambda_{\rm j}$ equals $|\overrightarrow{v_{\rm j}(z)} - \overrightarrow{v_\phi}(z)|$, where
$v_\phi= \omega z\sin(\theta_0)$ is the azimuthal velocity of the jets obeying to the 
precession rotation with the angular velocity $\omega=2\pi/P_0$ at the inclination $\theta_0$ 
to rotation axis at the distance $z$ from the source and with the precession period $P_0$.
In general, while the distance $z$ increases the vector of jet velocity $\overrightarrow{v_{\rm j}(z)}$
more and more deviates from the initial velocity vector --- hereinafter $Z$-axis of the 
frame of reference, which co-rotate with the jet, with $Y$-axis directed oppositely to
the precession movement, i.e. to vector $\overrightarrow{v_\phi}(z)$.

The brightness profile of the radio jets is approximated as
\beq
S_\nu(z)=S_{0} \exp(-z/\tau)
\label{Bright}
\eeq
with the brightness $S_{\nu} = 32.7$~mJy per a beam of telescope of a size 
$\phi_{\rm b}=0\arcsec32$ at $\nu = 4.86$~GHz and
at a distance $z = 50^{\rm d} \times v_{\rm j}^*$ (\cite[2010]{Rob10}; \cite[2011]{Bell11}), 
the decrement
$\tau = 55.\!\!^{\rm d}9 \times v_{\rm j}^*$ at flight times $t=50\div250^{\rm d}$
and $\tau = 115^{\rm d} \times v_{\rm j}^*$ at flight times $t>250^{\rm d}$, 
where $v_{\rm j}^* = 0.2647\,c$ is the fiducial speed.
Given the expression (\ref{Bright}) for the brightness
the differential brightness will be the function $s_\nu(z)=s_{0}\exp(-z/\tau)$,
with the normalization factor $s_{0} = S_{0}/2\tau \sinh(l_{\rm b}/2\tau)$, where 
$l_{\rm b}= \phi_{\rm b} D$ is the linear size of the telescope beam. The differential 
brightness in dependency on jet length is obtained as 
$s_\nu(l)=s_\nu(z) v_{\rm j\,z}(z)/\lambda_{\rm j}(z)$, 
where $v_{\rm j\,z}(z)$ is the $z$-component of jet velocity.

At flight times $t<50^{\rm d}$ the profile $S_\nu(z)$ is undetermined because of
contamination by the radio core. 
However, the exponential run of the brightness persists down to a flight time of $\sim 10^{\rm d}$
accordingly to the observations of higher resolution of \cite[(2008)]{Rob08}.
Transverse size of the jets at the plane of the sky is smaller than size of the beam,
hence the above definition of the differential brightness $s_\nu(z)$ of the jet remains legal
at flight times $< \phi_{\rm b} D/\theta_{\rm jr} v_{\rm j} \approx 290^{\rm d}$.
Extrapolation of $s_\nu(z)$ in the zone of radio brightening ($5.\!\!^{\rm d}4$) 
and estimation by formulae (\ref{press_m}, \ref{press_r}) gives a magnetic field $0.50$\,G,
whose difference with the result of \cite[(1987)]{Ver87}, $0.08$\,G, is
explained by accounting for the clumping of the jets.

The dynamic pressure (\ref{press_d}) on the surface of the quasi-continuous radio jets causes 
their acceleration. Now we have all constituents to write the equation of dynamics
\beq
\overrightarrow{a} = 
- \eta \frac{p_{\rm m} (1+\frac{1}{3 \beta_{\rm H}} + \beta_{\rm th}) \lambda_{\rm j} z \theta_{\rm jr}}{\dot{M}_{\rm j}}
\overrightarrow{e}_{\rm n}
\label{acc}
\eeq
of a jet segment, which is considered to be independent on other segments, of the length
$\lambda_{\rm j}$, of the transverse size $z \theta_{\rm jr}$, of the mass $\dot{M}_{\rm j}$, 
within the geometrical factor $\eta \sim 1$, depending on transverse profile of the jet. Here
$\overrightarrow{e}_{\rm n}$ is the unit normal vector to jet axis directed oppositely
to movement of the jet pattern. We believe that equation (\ref{acc}), in a fact 
an equation for the material point, correctly
describes the behavior of the precessing jet, which is really a quasi-continuous flow, 
while the displacements of the jet from the ballistic position is relatively small.

At angular distances beyond $\sim 4''$, or $t\sim 430^{\rm d}$, the SS\,433 radio jets look fragmentary 
(e.g. \cite[2010, Fig.~2]{Rob10}) --- likely the jets there are not constitute 
a continuous flow and the model is not true there. Another characteristic distance corresponds
to $t=250^{\rm d}$, where the slope of the brightness profile becomes flatter 
(\cite[2010]{Rob10}) and, on other hand, the angle of the impingement of swept gas and 
the jets becomes almost constant, namely normal to the jets.

\section{\large Simulation of dynamical jets of SS\,433}
The acceleration (\ref{acc}) was used to find locus of the SS\,433 radio jets.
The initial conditions of this kinematic problem are defined by the standard kinematic 
model, whose parameters, most accurate to now, are collected in Table~1. 
The simulated 3D locus was projected onto the plane of the sky accounting for the 
time delay of arrival of photons to an observer (the effect of light propagation), 
which differs for different points of the jet trajectory resulting in distortion
of jets image. 

\noindent
\vspace{0.3cm}
\begin{table*}[h]
\begin{center}
\caption{Parameters of standard kinematic model of SS\,433 jets}
\begin{tabular}{p{6cm}|p{4.0cm}|p{0.5cm}}
 \hline
                                       &                          & ref.$^{\rm 1}$  \\
 \hline
jet velocity & $v_{\rm j} = 0.2581\,c$      & [1] \\ 
 inclination of precession axis to the line of the sight & $i = 78\degree81$ & [1]\\
 precession cone half-angle            & $\theta_0 = 19\degree75$ & [1]\\
 precession period                     & $P_0 = 162.250$\,d       & [1]\\
 initial precession phase epoch$^{\rm 2}$                         & $t_0 =$\,JD\,2\,443\,508.41 & [1]\\ 
position angle of jet precession axis & $\chi = 98\degree2$      & [2] \\
\hline
\end{tabular}

\noindent $\mbox{ }^{\rm 1}$ [1] --- \cite[(2008)]{Dav08}, [2] --- \cite[(2002)]{St02}\\
$\mbox{ }^{\rm 2}$ of the minimal inclination of the east jet to the line of the sight\\
\end{center}
\end{table*}

\begin{figure}[ht]
\centerline{\hbox{\includegraphics[width=16cm]{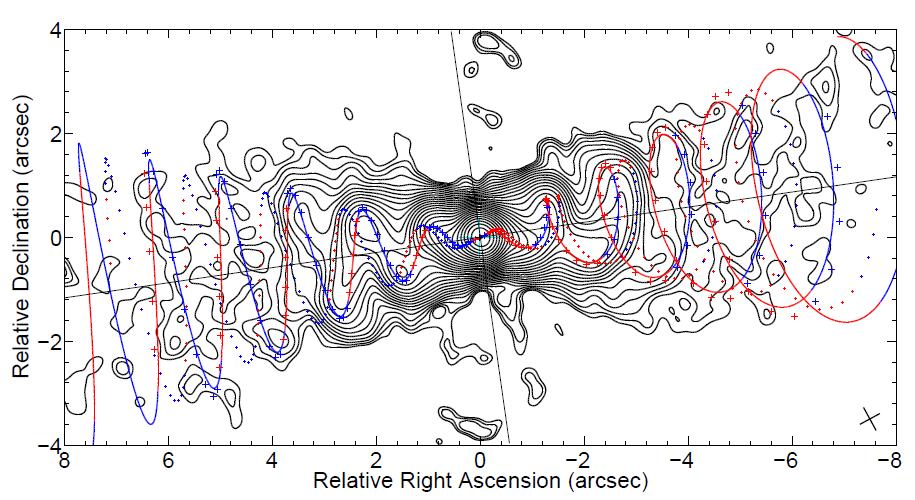}}}
\caption{
Simulated jets of SS\,433 for the pair $v_{\rm j} /D = 0.258\,c /4.8$\,kpc: the dynamic 
(pluses every $10^{\rm d}$ of flight) and ballistic (points per $5^{\rm d}$) jets, --- 
they are superimposed on the image of radio jets of SS\,433 taken from (\cite[2010]{Rob10}).
The solid line of the ballistic model jets for the pair $v_{\rm j} /D = 0.2647\,c /5.5$\,kpc
is of \cite[(2010)]{Rob10} and used as a template for fitting of the dynamical jets.
Approaching/receding parts of the jets are colored by blue/red.
A length per a simulated jet track is 800 flight days.
}
\label{F1}
\end{figure}

The model jets are shown in Fig.~\ref{F1}, where they are superimposed on the image of  radio 
jets of SS\,433 on July 11, 2003 taken from (\cite[2010]{Rob10}). This image is extremely 
deep and the jets are seen up to $800$ days of a flight.
It is seen that the dynamic jets, simulated in our model with the settings $v_{\rm j} /D = 0.2581\,c/4.8$\,kpc for the initial speed and distance, and the ballistic jets obtained by 
\cite[(2010)]{Rob10} 
with the settings $v_{\rm j} /D = 0.2647\,c/5.5$\,kpc, visually are almost indistinguishable in both
inner and external jets: the difference between the model jets is less than the accuracy
of determination of jet axis at the image. For other settings for the dynamic model of jets 
we chose following characteristic values:
a geometry factor $\eta = 1$; a ratio of thickness of the radio bright shell around a jet cloud to 
size of the cloud $l_{\rm sh}/l_{\rm cl}=0.5$, the finding of \cite[(2012)]{Ino12} for
supernova remnants; a pressure ratio of thermal gas and magnetic field $\beta_{\rm g} = 1$, 
that is expected at the vicinity of jets clouds where the magnetic field is amplified
(\cite[1996]{Jon96}; \cite[2012]{Ino12}); a ratio of energy densities of the magnetic field
and relativistic particles $\beta_{\rm H} = 3/4$, that corresponds to minimal total energy
of the field and particles (\cite[1979]{Gin79}); an initial flux of kinetic energy in the jet
contained in the clouds ${L}_{\rm j} = 10^{39}$\,erg/s and a temperature of the jet
clouds $T_{\rm cl} = 2\cdot 10^4$~K at a distance of $10^{15}$\,cm, these are 
approximately determined by observations of the optical jet. The rate of jet mass flux 
$\dot{M}_{\rm j}$ was supposed to be constant through jet length. Only two parameters 
were fitted: an exponent of the clouds temperature profile
$n=1.50^{+0.07}_{-0.05}$, that influences
on synchronism of the fits to inner and extended jets; and a ratio of energies of all 
relativistic particles and relativistic electrons $\beta_{\rm e} = 2.7 \pm 0.4$, 
although it is not an independent fitting parameter.
The accuracies of the fitted parameters correspond to a deviation of the dynamical model
jet from the ballistic model jet of a half of an image resolution $0\arcsec47$ 
at a distance of $6"$, on which the position of precession spiral of eastern jet is still clearly discernible.
The dynamics of the radio jets was taken into account only in the region from the zone 
radio brightening to a distance $t_1 =215^{\rm d}$ of a flight, where the filling
factor $f$ is close to $1$ and the model becomes invalid.

%
%

\begin{figure}[ht]
\centerline{\hbox{\includegraphics[width=14cm]{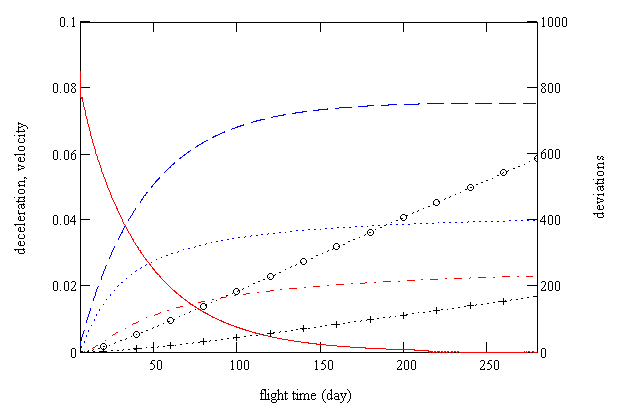}}}
\caption{
Kinematics deviations of the dynamical jet from the ballistic jet in dependency on flight time.
Along left axis: tangential deceleration, $\dot{v}_{\rm j}$, in units of $c/P_0$ --- solid red line; 
relative decrement of velocity, $1-v_{\rm j}(t)/v_{\rm j}(0)$ --- dashed blue line.
Along right axis: longitudinal, $-\Delta z$, and transverse, $\Delta y$, deviations of the jet 
locus in the comoving 
frame of reference, in units of mas --- dotted black lines with plus and circle symbols, respectively;
increment of azimuth of precession rotation, $\Delta \psi$, in units of degree$/10$ --- dotted blue line;
increment of precession cone semi-opening, $\Delta \theta_0$, in units of degree$/100$ --- 
dash-dotted red line.
}
\label{F2}
\end{figure}

Deceleration and deviation of kinematics of the dynamic jets, with the parameters given above, 
from the ballistic ones is shown in Fig.~\ref{F2}. The profile of the deceleration $a(t)$, that is
module of the tangential component of the jet acceleration vector relative to the jet velocity vector, 
shows that the deceleration occurs mainly in the inner jets and peaks at $0.082\,c/P_0$ 
in the zone of radio brightening, as \cite[(2004)]{St04} have guessed. An average deceleration 
in the first precession cycle is $0.0191\,c/P_0$. A relative jet speed decrement 
is $\delta v_{\rm j} = 7.5\%$, a half of it is accumulated before a distance of 
$t_{1/2} = 32.\!\!^{\rm d}6$, i.e. for one-fifth of the precession period.
Approximately the same distance limits the zone of significant slowdown of the jets: 
the deceleration beyond this distance leads to a shift of the precession spiral by a half 
of the image resolution $\phi_{\rm b}/2$ at a distance of approx $6"$. Thus, 
the external jet, at $t > P_0$, should be like ballistic. During a flight time $P_0$, 
or in bounds of $1\arcsec386$, the precession cone semi-opening increases by
$\Delta \theta_0 = 2\degree0$, the azimuth relative to the axis of precession increases 
(or the precession phase decreases) by $\Delta \psi = 37\degree7$, or the jet offsets 
along the azimuth by $\Delta y =323$~milliarcsec (mas) in the direction opposite to the precession 
rotation direction, which is approximately 4 times more than the radial offset 
$|\Delta z| = 85$~mas against $Z$-axis --- the precession helix becomes more 
compressed and twisted. These should show up as slowing down and precession lag 
of the jets relatively to the standard kinematic model. An average visual slowdown --- 
along $Z$-axis, which is observable as compression of the helix --- is $0.0423\,c/P_0$ 
in the first precession cycle and $0.0230\,c/P_0$ during a flight time of $350^{\rm d}$.
The latter is close to $0.02\,c/P_0$, the value found by \cite[(2004)]{St04} approximately at
the same length of a flight. They also pay attention to the delay of the jets in precession
phase. The fact of a significant offset $\Delta y$ of the model dynamic jets allows to explain the
appearance of individual clouds outside the SS\,433 jets track as a result of uneven 
dynamic pressure of inhomogeneous wind on the fragmented radio jets.

%
%

\section{\large Discussion}
We found that the SS\,433 jets should be noticeably decelerating to satisfy to
observed radio synchrotron brightness, if the radiating relativistic electrons are 
steadily injected by shocks in the jets. In the colliding with ambient medium
the jet acquires the additional momentum which lies in the tangential plane to the precession cone
and makes the jet to decelerate along $Z$-axis and shift along $Y$-axis, in the reverse 
direction to the precession. The acquired shift must be observed as an increase of torsion of
the precession helix, or a decrease of the precession phase. Note that the known azimuthal 
asymmetry of radiation of the optical jets points on a similar character 
of the interaction (\cite[1997]{Pan97}). In the case of only head collision of clouds 
of a jet with impinging gas would be observed only deceleration of the jet.

The dynamical model of the jets does good fit to observed jets of SS\,433 at a distance of $ 4.8$\,kpc.
The deceleration has maximum in the zone of radio brightening and decreases further. 
So the maximum of radio jets radiation associates with the maximum of the jets kinetic energy
dissipation.
The zone of deceleration, i.e. where the deceleration is essential for the jets, spans 
flight time of only one-fifth of the precession period. Beyond the zone of deceleration
the dynamical jets imitate ballistic ones with an initial velocity 
$0.2647\,c$ and at a higher distance to the object $5.5$\,kpc.
There are contradictory estimations of the distance, including
those obtained by the kinematic method on the basis of observations of the inner and
extended radio jets. The dynamical model soothes the dichotomy of the kinematic distance.

\cite[(2004)]{BB04} have fitted by the kinematic model, allowing the jets speed to
variate stochastically, the extended radio jets, i.e. on the scales where the jets
are already ballistic in our dynamical model. So, jets deceleration does not
influence on their conclusions. A speed variations amplitude they found is of
$0.032 (4.8\,{\rm kpc} /5.5\,{\rm kpc}) = 0.028\,c$, or $12\%$ of the jets speed, 
within parameters of our model, that is comparable with a speed relative decrement
$\delta  v_{\rm j} = 7.5\%$ originated from the jets deceleration.
These speed variations may be due partly to the uneven process of jets slowing.
The results of \cite[(2004)]{BB04} do not exclude the dynamic model, which shows
the extended jets are ballistic ones, with parameters $v_{\rm j}= 0.2386\,c$ and 
$D= 4.8$\,kpc, being an intermediate to the two options they considered. 

It is quite possible the blurring of the nutation structure of the jets within the zone of 
deceleration, because the ram pressure on the jets would be strongly modulated
with nutation phase. Besides, the blurring could be initiated by abrupt heat and expansion of
the jet clouds and decollimation of the jets in the zone of radio brightening, where
pressure in the clouds should fall (\cite[1993]{Ver93}; \cite[1999]{Pan99}). 

Found above physical parameters of the dynamical model are not independent of each other.
They allow variations which leave the acceleration to be invariant:
\beq
a(z) = k_{\rm a}(z) \left(\frac{1+\frac{1}{3 \beta_{\rm H}} + \beta_{\rm g}}{\dot{M}_{\rm j}}\right)^{7/3}
\left({\frac{\beta_{\rm H} \beta_{\rm e}}{k_{\rm V} T_0}}\right)^{4/3},
\label{acc_m}
\eeq
as obtained from (\ref{press_r}) and (\ref{acc}), where $k_{\rm a}(z) = 
\eta z^{1-4n/3} \theta_{\rm jr}  \lambda_{\rm j}^{7/3} 
(k_\nu \mu m_{\rm p} D^2 s_\nu/ k_{\rm B} )^{4/3}$.
Nevertheless, the found magnitudes of $l_{\rm sh}/l_{\rm cl}$, $\beta_{\rm H}$, 
$\beta_{\rm e}$, $\beta_{\rm g}$, $\dot{M}_{\rm j}$ and $T_0$ are 
observationally and theoretically justified with an accuracy of a some. 

A power of loss of the radio jet kinetic energy 
$\dot{W}_{\rm k} \approx 2 \delta v_{\rm j} {L}_{\rm j} = 1.50\cdot 10^{38}$~erg/s
is huge and unobserved radiatively. Possible, this power is drained into momentum of the wind
colliding with the jets and transformed eventually into momentum of W\,50
(\cite[1998]{Dub98}). A high efficiency of the mechanical energy transfer from jets
to atomic and molecular outflows in the surroundings is observed also in AGNs 
(\cite[2013]{Mor13}). 

The dynamical model of the jets satisfying the jets kinematics has implications on
physics of the jets. The filling of the clumped jets increase with distance due to
clouds heating and expansion, while clouds pressure is determined by the ram 
pressure on the jets. The clouds temperature rises from $1.4\cdot 10^5$~K in the
brightening zone to $3\cdot 10^7$~K at flight times $\sim t_1 = 215^{\rm d}$ where the
filling $f \rightarrow 1$ and the model becomes invalid. Really, the model jets kinematics in not sensitive
to departures of jets physics from the dynamical model beyond the deceleration
zone. Possibly, the clouds reheating is non-monotonic. Given the temperature increment
$\Delta T_{\rm cl}$ a power of jet heating is
$Q=k_{\rm B}\Delta T_{\rm cl} (3\dot{M}_{\rm j}/2\mu m_{\rm p}) = 
2.1 \cdot 10^{35}$~erg/s, that is far more than an observed jet X-ray luminosity of
$\sim 10^{33}$~erg/s and surprisingly close to the radio jet synchrotron luminosity.

Our work shows that there is a small deceleration of the radio jets and, on arcsecond-scales, they 
can be imitated by ballistic jets choosing inappropriate distance to the object. Explicit
detection of the deceleration is a problem for further observations of the inner radio jets.

\end{document}